\PassOptionsToPackage{dvipsnames}{xcolor}
\documentclass[twocolumn]{aastex631}  

\usepackage{lipsum}
\usepackage{adjustbox}
\usepackage{xcolor}
\usepackage{footmisc}
\usepackage{hyperref}

\newcommand{\viewer}{\texttt{Viewer}}

\shorttitle{The SNAD ZTF viewer}
\shortauthors{Malanchev et al.}

\begin{document}

\title{The SNAD Viewer: Everything You Want to Know about Your Favorite ZTF Object}

\author[0000-0001-7179-7406]{Konstantin Malanchev}
\affil{Department of Astronomy, University of Illinois at Urbana-Champaign, 1002 W Green Street, Urbana, IL 61801, USA}
\affil{Sternberg Astronomical Institute, Lomonosov Moscow State University, Universitetsky pr. 13, Moscow 119234, Russia}
\author{Matwey V. Kornilov}
\affil{Sternberg Astronomical Institute, Lomonosov Moscow State University, Universitetsky pr. 13, Moscow 119234, Russia}
\affil{National Research University Higher School of Economics, 21/4 Staraya Basmannaya Ulitsa, Moscow, 105066, Russia}
\author[0000-0001-7178-0823]{Maria V. Pruzhinskaya}
\affil{Sternberg Astronomical Institute, Lomonosov Moscow State University, Universitetsky pr. 13, Moscow 119234, Russia}
\author[0000-0002-0406-076X]{Emille E. O. Ishida}
    \affil{Universit\'e Clermont Auvergne, CNRS/IN2P3, LPC, F-63000 Clermont-Ferrand, France}
\author[0000-0002-6298-1663]{Patrick D. Aleo}
\affil{Department of Astronomy, University of Illinois at Urbana-Champaign, 1002 W Green Street, Urbana, IL 61801, USA}
\affil{Center for AstroPhysical Surveys, National Center for Supercomputing Applications, Urbana, IL, 61801, USA}
\author{Vladimir S. Korolev}
\affil{Independent Researcher}
\author{Anastasia Lavrukhina}
\affil{Faculty of Space Research, Lomonosov Moscow State University, Leninsky Gori 1 bld. 52, Moscow 119234, Russia}
\author{Etienne Russeil}
\affil{Universit\'e Clermont Auvergne, CNRS/IN2P3, LPC, F-63000 Clermont-Ferrand, France}
\author[0000-0002-6423-1348]{Sreevarsha Sreejith}
\affil{Physics Department, Brookhaven National Laboratory, Upton, NY 11973}
\author{Alina A. Volnova}
\affil{Space Research Institute of the Russian Academy of Sciences (IKI), 84/32 Profsoyuznaya Street, Moscow, 117997, Russia}
\author{Anastasiya Voloshina}
\affil{Universit\'e Clermont Auvergne, CNRS/IN2P3, LPC, F-63000 Clermont-Ferrand, France}
\author[0000-0002-2308-6623]{Alberto Krone-Martins}
\affiliation{Donald Bren School of Information and Computer Sciences, University of California, Irvine, CA 92697, USA}
\affiliation{CENTRA/SIM, Faculdade de Ci\^{e}ncias, Universidade de Lisboa, Ed. C8, Campo Grande, 1749-016, Lisboa, Portugal}


\begin{abstract}
We describe the SNAD \viewer, a web portal for astronomers which presents a centralized view of individual objects from the Zwicky Transient Facility's (ZTF) data releases, including data gathered from multiple publicly available astronomical archives and data sources.
Initially built to enable efficient expert feedback in the context of adaptive machine learning applications, it has evolved into a full-fledged community asset that centralizes public information and provides a multi-dimensional view of ZTF sources.
For users, we provide detailed descriptions of the data sources and choices underlying the information displayed in the portal.
For developers, we describe our architectural choices and their consequences such that our experience can help others engaged in similar endeavors or in adapting our publicly released code to their requirements.
The infrastructure we describe here is scalable and flexible and can be personalized and used by other surveys and for other science goals.
The \viewer\ has been instrumental in highlighting the crucial roles domain experts retain in the era of big data in astronomy.
Given the arrival of the upcoming generation of large-scale surveys, we believe similar systems will be paramount in enabling an optimal exploitation of the scientific potential enclosed in current terabyte and future petabyte-scale data sets.
The \viewer\ is publicly available online at \url{https://ztf.snad.space}.
\end{abstract}





\section{Introduction}
\label{sec:introduction}


Modern astronomical surveys have drastically changed the process of astronomical data analysis.
Traditionally, the field of astronomy was based on small data sets. Each of them gathered, analyzed, and reported by small research groups.
Recently, the situation has drastically transformed into big data volumes, involving a large number of individuals at each stage of the process: from data ingestion to the final publication of scientific results.
This long-predicted new paradigm \citep[e.g.,][]{2001Sci...293.2037S} fostered the development of crafted infrastructure within each survey, including strategies for storage, indexing, and archiving protocols specifically designed to fulfill the needs of a given scientific goal or community.
The resulting data environments have enabled the successful application of machine learning techniques in astronomical data, especially for classification and regression tasks -- two of the most well-known examples being photometric classification of supernovae \citep[e.g.,][]{Jones2018, vicenzi2022} and photometric redshift estimation \citep[e.g.,][]{zhou2021, abbott2022}.
For such supervised learning tasks, once the experiment design and analysis pipeline is finalized, researchers can rerun the entire machinery within the same survey, updating results as more data become available, in principle, without any need for visual screening of the original data.

At the same time, the observational nature of astronomy, coupled with the predetermined scanning strategies of large surveys, results in a tremendous potential for discovery~\citep[see, e.g.,  LSST-related reviews][]{lsst-white-book,lsst-tvs-roadmap}.
In this context, human intervention is unavoidable.
So far, even the most efficient machine learning algorithm cannot take into account all available data about each object it considers and can only provide good candidates as its output.
These candidates usually need to be scrutinized by an expert who will put each of them in context and give meaning to a new discovery \citep{dick_2013}. The expert, in turn, requires as much information as possible to form a comprehensive view of the new candidate and ensure its significance, as well as its novelty, reaching a final human conclusion.
This means, for example, the researcher must gather data
about the same object from multiple sources, build correlations
between various wavelengths, and search for similar matches
among well-known objects.
In the specific case of transients or variables, one will also be interested in all available legacy data that can provide clues about their time evolution. 

From the infrastructural point of view, this imposes a new set of requirements, which includes enabling easy visualization and cross-match between databases from different surveys, as well as with modern static catalogs, historical data, and derived data products. The SNAD team\footnote{\url{https://snad.space}} met this challenge when our experts aimed to analyse a large set of variable objects.
That is the reason why The SNAD ZTF viewer was originally built (hereafter, the \viewer): the \viewer\ was designed and built to optimize the allocation of human resources in astronomical investigative tasks.
It helps experts make decisions about the class and properties of a given object faster, consolidating photometry, cross-matching, and other information about astronomical sources retrieved from multiple datasets and data archives, on a single web page. Moreover, the \viewer\ is publicly openly available online at \url{https://ztf.snad.space}.

The time spent by an expert trying to mine information about an object became crucial when we started using active anomaly detection algorithms \citep{Ishida_etal2021}. These human-in-the-loop learning strategies use feedback from a domain expert to guide adaptations of hyperparameters of a traditional machine learning model, thus constructing a model tailored to the expert's own anomaly definition. The issue of scarce human resources for manual screening was already a limiting factor in the first stages of the SNAD pipeline development when we were dealing with a few thousand objects from the Open Supernova Catalog~\citep{pruzhinskaya_etal2019}. However, it  became unmanageable when 
faced with the Zwicky Transient Facility data releases\footnote{\url{https://www.ztf.caltech.edu/ztf-public-releases.html}} (ZTF DR). 

The Zwicky Transient Facility (ZTF) is a  time-domain photometric survey currently in progress using the Palomar 48-inch Schmidt telescope~\citep{Bellm_etal2019a,Bellm_etal2019b}. It represents the state of the art of large-scale astronomical surveys and is considered a precursor to the upcoming Vera C.~Rubin Observatory Legacy Survey of Space and Time\footnote{\url{https://www.lsst.org}\label{foot:lsst}} (LSST). For instance, data release 13, contains 4.37 billion light curves constructed from hundreds of billions of single-exposure extractions\footnote{\url{https://www.ztf.caltech.edu/ztf-public-releases.html}}.

There are a few publicly available tools that enable different levels of interaction with ZTF data, chief among them being the portals from ZTF community brokers (Section \ref{sec:ztf-drs}), IRSA IPAC\footnote{\url{https://irsa.ipac.caltech.edu}\label{foot:irsa}}, \added{ and Fritz Astronomy Marshal\footnote{\url{https://fritz.science}\label{foot:fritz}}}.
Although they share some common features with the \viewer, broker systems were designed to deal with the data from the alert stream and to enable live analysis, while our goal was to explore the ZTF DRs. IRSA IPAC, on the other hand, provides a web user interface to access photometric data of surveys conducted by many different missions.
Despite providing a convenient framework to handle ZTF DRs, its scope was not broad enough to fulfill the requirements of the SNAD anomaly detection pipeline.
For expert analysis we required a portal providing instant access not only to a single object's light curve, but also to as much relevant contextual information as possible, including nearby ZTF DR objects and cross-matched results from different catalogs.
Moreover, IRSA IPAC gives access to the most recent data releases, while for the SNAD team, it was crucial to support a set of legacy data releases, whose data might be involved in ongoing projects.
\added{
Fritz Astronomy Marshal is a ZTF platform produced by the project's team and combining data from data releases, alert stream and cross-matching information from many catalogs.
Access to the portal is restricted, but its source code is open and easy to run. 
However there is no easy way to ingest data-release photometry into the self-hosted instance of the system.
}
Among non-ZTF frameworks with similar goals and infrastructure, we highlight the Open Supernova Catalog \citep[OSC, ][]{osc}, the OGLE survey portal\footnote{\url{http://ogle.astrouw.edu.pl}}~\citep{ogle3,ogle4}, and the ASAS-SN portal\footnote{\url{https://asas-sn.osu.edu}}~\citep{asassn_server,asas-sn3}, all of which are also integrated into our \viewer. 

So far, the \viewer\ enabled all the ZTF-based results reported by the SNAD team \citep{Malanchev_etal2021, Pruzhinskaya_etal2022, Aleo_etal2022}, whose summary can be found in the SNAD catalog\footnote{\url{https://snad.space/catalog/}\label{foot:cat}}, currently hosting 144 candidate transients found by the SNAD team on ZTF DRs, while being absent from systematic searches from other groups.

Once it was made publicly available, the \viewer\ made its way from being an infrastructure project used exclusively by SNAD experts to becoming a valuable community resource for scrutinizing ZTF DR objects. It is currently being integrated into two ZTF brokers, ANTARES\footnote{ \url{https://antares.noirlab.edu}\label{foot:antares}}, \citep{antares}, and Fink\footnote{\url{https://fink-broker.org}\label{foot:fink}} \citep{fink} and also in the Young Supernova Experiment marshal \citep{yse, Coulter2022_YSEPZ}. It also counts on average, a few dozen unique visitors per day from multiple countries. Beyond ZTF DRs, it provides access to the ZTF alert photometry and light curves from Pan-STARRS \citep{panstarrs} and Gaia \citep{2016A&A...595A...1G, gaiadr3} surveys. It is an ideal entry point for the photometric investigation of various types of variable objects, including active galactic nuclei (AGN), Milky Way variable stars, and microlensing events. 

In what follows, we give details on the currently available services and their underlying implementation. Details about ZTF DRs are given in Section \ref{sec:ztf-drs}. Section \ref{sec:viewer} notifies users about the displayed information available and the choices which lead to them. Section \ref{sec:akb} describes the only part of the \viewer\ that is, for the moment, private, where SNAD experts can annotate individual objects for subsequent internal use. Sections \ref{sec:infrastructure} and \ref{sec:impl} describe details of our implementation and is directed toward developers and researchers who may find our experience useful in the construction of similar systems. We present our conclusions and plans for further developments in Section \ref{sec:conclusions}.

\section{The Zwicky Transient Facility Data Releases}\label{sec:ztf-drs}

ZTF observes all of the  visible Northern sky, covering from 25,000 to 30,000~square degrees having a field of view of 47~square degrees in $gri$ passbands, and reaching a limiting median $r$-magnitude of $\sim20.6$ for a typical exposure of 30 seconds.
It runs both public and private surveys with different cadence, exposure, and passband usage.
The public survey is the source of the photometric alerts sent to community brokers, including ALeRCE\footnote{\url{https://alerce.science}}~\citep{alerce}, AMPEL\footnote{\url{https://github.com/AmpelProject}}~\citep{ampel}, ANTARES\footref{foot:antares}~\citep{antares}, Fink\footref{foot:fink}~\citep{fink}, Lasair\footnote{\url{https://lasair-ztf.lsst.ac.uk}} and MARS\footnote{\url{https://mars.lco.global}}.
It operates using a differential photometry pipeline that triggers alerts using current observations and the ZTF reference catalogs.
The survey has had two phases so far.
During Phase~I (March 2018 -- September 2020) it had a 3-day cadence for extra-galactic fields and a 1-day cadence for Galactic fields, while in Phase~II (from December 2020 onwards) it switched to a homogeneous 2-day cadence all over the observable sky.

The public survey, which is openly available, uses 30-second exposures and primarily operates in $gr$ passbands.
The private survey used 60\% of its observational time during Phase~I and 50\% during Phase~II. 
It includes a significantly higher fraction of $i$-passband observations and also high-cadence data reaching hundreds of consecutive observations of the same field during a single night.

ZTF DRs were announced every six months from DR~1 (May 2019) to DR~4 and then the releases switched to a bimonthly schedule.
Each DR covers observations for both private (from the start of the survey up to 18 months before the release date due to a proprietary period) and public surveys (from the start of the survey up to a few weeks before the release date).
The ZTF DR photometric pipeline is different from the alert pipeline -- it is based on source extraction for individual frames and subsequent cross-matching of these sources along all frames within a single pair of observation field/CCD quadrant and passband.
This leads to two peculiarities of the data-release objects compared to the alert stream: 1) object light curves do not include non-detections \added{\footnote{The alert stream has a $5\sigma$ detection threshold, but it also provides historical measurements up to 30-days before each detection.}}, 2) a single sky source can be represented by more than a \textit{dozen} ZTF DR objects due to three passbands and overlapping observation fields.

There are other important differences between the DRs and the alert stream: for example, source identifiers are independent and have different formats, the data-release pipeline doesn't contain the bogus-to-real image classification step which is a part of the alert production process~\citep{Duev_etal2019}, DRs provide Heliocentric MJD (HMJD) time of the middle of exposure while alerts use exposure start JD, among others.
Alerts, in turn, are based on source detection via difference imaging,  which is more effective for extragalactic sources.
For example, most of the ZTF Bright Transient Survey supernovae~\citep{2020ApJ...895...32F,2020ApJ...904...35P} are missing from the DRs.
However, some robust SN candidates not presented in the alert stream were found in ZTF~DRs by the SNAD team in~\citet{Aleo_etal2022,Pruzhinskaya_etal2022}.
Moreover, the DR light curves have better limiting magnitudes and may also include observations that do not pass the difference imaging pipeline  detection threshold (see Section~\ref{sec:antares-lc} and Fig.~\ref{fig:antares}).
\added{To address these issues IPAC provides a service for PSF-fit forced-photometry\footnote{\url{https://web.ipac.caltech.edu/staff/fmasci/ztf/forcedphot.pdf}\label{foot:forced-phot}}, which has both deeper limiting magnitudes and includes non-detections, but can only be used for individual sky positions.}

IRSA IPAC\footref{foot:irsa} provides different ways to access ZTF DRs: through the web interface, light-curve API calls, and via bulk-downloadable files.
Since the first two options are not scalable to access the dozens of millions of light curves needed for the SNAD anomaly detection pipelines,
we decided to use bulk-downloadable files, which we converted to our internal database format (further described in Section~\ref{sec:data-storage}).
However, this option does not provide as much data as the other two interfaces, limiting object properties to: average coordinates, field, readout-channel identifiers, and passband. 
Similarly,  detection properties are restricted to Heliocentric MJD of the middle of the exposure, the magnitude and its uncertainty, the color correction coefficient, and the quality flags.

It is difficult to overstate the importance of ZTF DR for time-domain astronomy. It has already motivated searches and studies of
AGNs \citep[e.g. ][]{sanchez2021},
strongly lensed QSOs \citep[e.g.][]{2021ApJ...921...42S}, 
microlensing \citep[e.g. ][]{rodrigues2022},
variable stars \citep[e.g. ][]{ztf-periodic, kupfer2021},
young stellar objects \citep[e.g.][]{spicy},
eclipsing binary systems \citep[e.g. ][]{kosakowski2022}
and, unexpectedly, supernova-like transients \citep{Pruzhinskaya_etal2022}.
Moreover, their volume and complexity, combined with their timely existence as a precursor to LSST, makes ZTF DR a unique ground for preparing data mining and machine learning techniques \citep{Malanchev_etal2021, Aleo_etal2022} which will be of paramount importance for the next generation of telescopes and surveys.
We did our best to make the \viewer\ a convenient place to inspect photometric and complementary ZTF DR data, a web portal that can serve all researchers interested in deep investigation of individual objects.

\section{The SNAD ZTF viewer web-portal}
\label{sec:viewer}

\subsection{Homepage}\label{sec:homepage}

\begin{figure*}
    \centering
    \includegraphics[width=\textwidth]{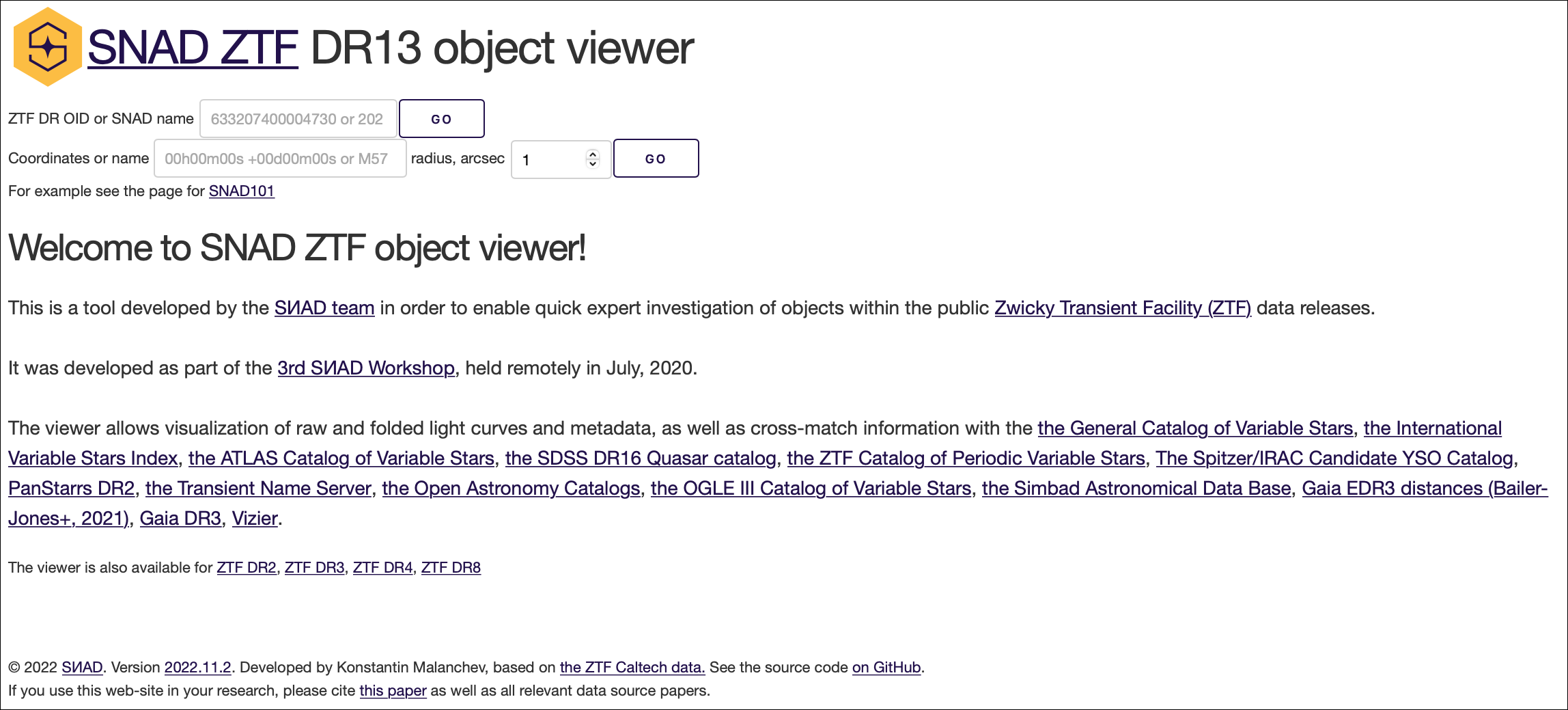}
    \caption{SNAD \viewer\ homepage, version 2022.11.2}
    \label{fig:homepage}
\end{figure*}

The \viewer\ homepage, shown in Fig.~\ref{fig:homepage}, gives a brief description of the website and references to the data sources available through the portal. 
Its header contains a login link (see Section~\ref{sec:akb} for SNAD internal resources description) and two search fields: one for ZTF object identifier (OID) and another one for cone search. The OID search field supports 15-16 digit strings for the ZTF DR identifier and 3-digit strings for the SNAD catalog objects\footref{foot:cat}.
The cone search field accepts an equatorial coordinate string in various formats\footnote{The \texttt{astropy.coordinates.SkyCoord} class resolves the coordinates, thus supporting most common formats of coordinate representations \citep{astropy}.} as well as object identifiers which are resolved to sky positions using SIMBAD~\citep{simbad}.
The cone search radius is to be specified in arcsecond -- the default value is 1\arcsec and the maximum supported radius currently is 60\arcsec  .

\subsection{Cone-search results page}\label{sec:cone-search-page}

\begin{figure*}
    \centering
    \includegraphics[width=\textwidth]{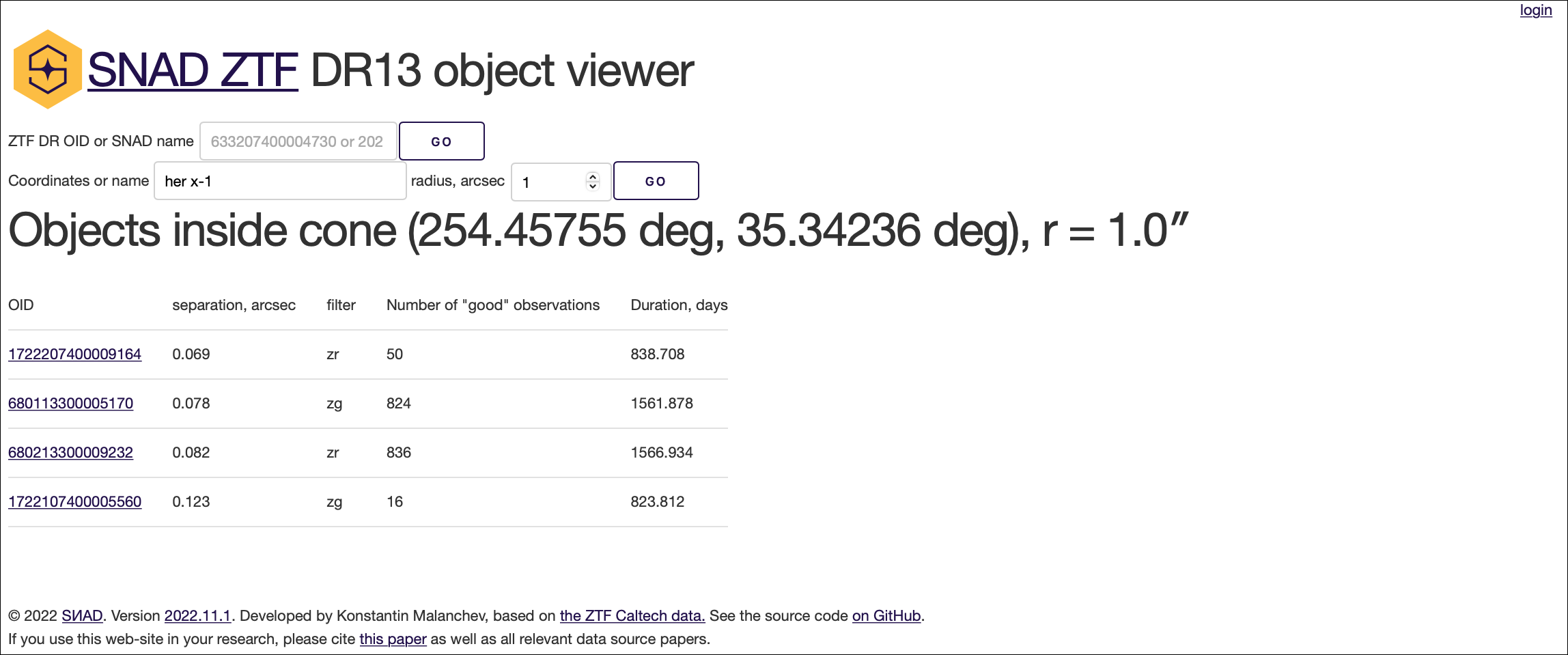}    \caption{Cone-search result page for a request ``her x-1'' and $1\arcsec$ cone radius.}
    \label{fig:cone-search-page}
\end{figure*}

After the user requests a cone search, they are redirected to a page displaying the table of search results.
This page has three states: (1) requested string cannot be parsed or resolved via SIMBAD, (2) no results are found, or (3) the search is successful and a table of matches is shown.
For the last case, resolved coordinates of the sky position are shown as well as all matched ZTF OIDs. An example is shown in Fig.~\ref{fig:cone-search-page}.
We note that no crossmatch to ZTF objects is performed at this step. Thus, due to the ZTF DR object definition, the user may see multiple object identifiers corresponding to a single sky source. After selecting one of the results, the user is redirected to the object page.

\subsection{ZTF object page}\label{sec:oid-page}

\begin{figure*}
    \centering
    \includegraphics[width=\textwidth]{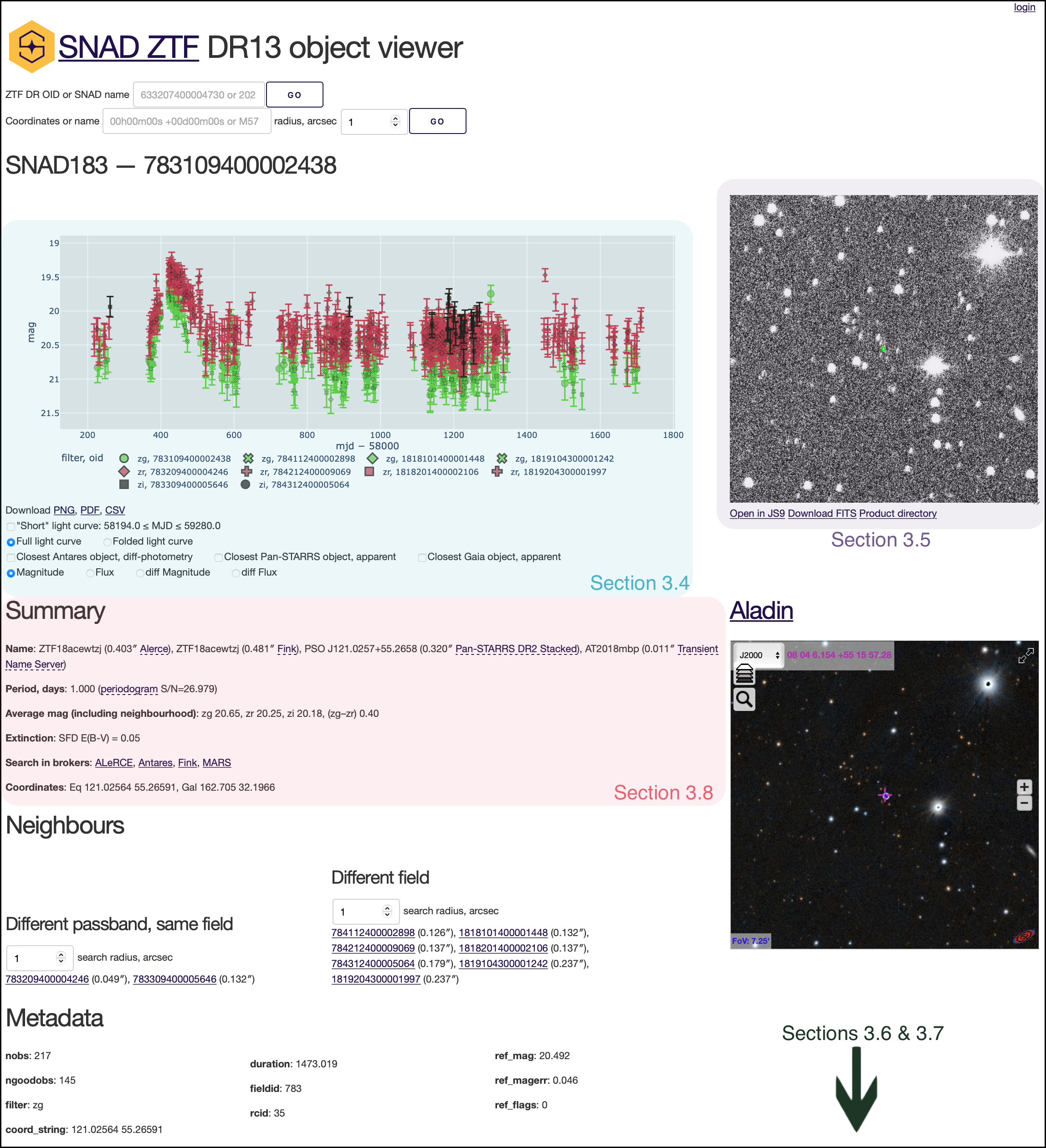}
    \caption{The upper part of the object page for SN candidate \citep{Pruzhinskaya_etal2022}  OID~\texttt{783109400002438} / SNAD183 / AT2018mbp. Light-curve plot (top left) shows object's $g$-passband photometry by large green circles, while the smaller symbols show photometry of eight more ZTF DR objects found within one arcsecond from its position in $gri$ passbands and four overlapping ZTF fields. The horizontal axis label, ``mjd'', denotes Heliocentric MJD at the middle of the exposure, as reported in ZTF DR. The FITS viewer (top right) shows a scientific image for a detection near the peak selected by a user. Aladin~\citep{aladin,aladin-lite} Sky Atlas (below the FITS image) shows the corresponding image from Pan-STARRS~\citep{panstarrs} while the small blue circle indicates the object position.}
    \label{fig:oid-page}
\end{figure*}

Fig.~\ref{fig:oid-page} shows the upper part of the object page, the lower part contains external (non-ZTF) catalog cross-matches (see Section~\ref{sec:external-catalogs}), light-curve features (see Section~\ref{sec:light-curve-features}) and the ZTF DR photometry table.
The ZTF object page is built from multiple blocks, and we describe the most important ones in the next subsections.

\subsection{Light-curve plot}\label{sec:light-curve-plot}

The light-curve plot is the main element of the object page.
By default, it shows ZTF DR light curves for the given object and all neighboring objects found within the cone of the default search radius ($1\arcsec$).
Data points from the target object have larger markers, so that it is easier to distinguish them visually from those of nearby objects.
This representation is configurable, and the user can hide light curves of arbitrary objects and change the search radius.
It is also possible to display a ``short" light curve, restricted to dates corresponding to those of the available private survey.
This allows visualization of a homogeneous cadence and overall survey properties presented in ZTF observations.
The SNAD team used this functionality in~\cite{Malanchev_etal2021}, \cite{Pruzhinskaya_etal2022}, and \cite{Aleo_etal2022}.

A period-folded representation of light curves is also available, see an example in Fig.~\ref{fig:gaia-lc}.
By default, it uses a period corresponding to the highest Lomb--Scargle periodogram peak \citep{Lomb1976,Scargle1982} calculated for the target ZTF DR object (Section~\ref{sec:light-curve-features}) and zero phase with respect to $\mathrm{HMJD} = 58000$.
The user can change both phase and period, which can possibly be obtained from multiple sources listed in the ``Summary'' section just below the plot (see Section~\ref{sec:summary-section}).

Both the original and the period-folded light curves can be downloaded in PNG or PDF formats, generated using the \texttt{matplotlib} library~\citep{matplotlib}.
These plots display the ZTF object neighbors and the ZTF range and can also handle user-configured parameters such as custom titles and additional photometric data.

Another available customization for the representation of photometric data is the choice between magnitude and flux and between their apparent and difference values.
Since ZTF DRs provide the photometric data in magnitudes, we transformed them into fluxes using the AB-system zero-point,
\begin{equation}\label{eq:flux_apparent}
    f = 10^{-0.4 (m - 8.9)} \,\mathrm{Jy} \,,
\end{equation}
\begin{equation}\label{eq:flux_err_apparent}
    \sigma_f = 0.4 \ln{(10)} \cdot f \cdot \sigma_m \,,
\end{equation}
where $m$ and $\sigma_m$ are the apparent magnitude and its uncertainty as reported in the ZTF DR, $f$ and $\sigma_f$ is our estimate of the apparent spectral flux density and its uncertainty in janskies.
The ZTF reference catalog (see Section~\ref{sec:fits-proxy} for the technical details) provides the reference magnitude and corresponding uncertainty for each OID.
When a user selects the differential photometry option, we load these values for each object and use them for differential flux density estimation,
\begin{equation}\label{eq:flux_diff}
    \Delta f = f - f_\mathrm{ref} \,,
\end{equation}
\begin{equation}\label{eq:flux_err_diff}
    \sigma_{\Delta f} = \sqrt{\sigma_f^2 + \sigma_{f_\mathrm{ref}}^2} \,,
\end{equation}
where $f_\mathrm{ref}$ and $\sigma_{f_\mathrm{ref}}$ are the reference flux density and its uncertainty determined by equations \ref{eq:flux_apparent} and \ref{eq:flux_err_apparent}.
Difference magnitudes are considered to have asymmetric uncertainties,
\begin{equation}\label{eq:mag_diff}
    \Delta m = 8.9 - 2.5 \lg{\Delta f} \,,
\end{equation}
\begin{equation}\label{eq:mag_err_plus_diff}
    \sigma_{\Delta m}^+ = -2.5 \lg{\left( 1 - \frac{\sigma_{\Delta f}}{\Delta f} \right)} \,,
\end{equation}
\begin{equation}\label{eq:mag_err_minus_diff}
    \sigma_{\Delta m}^- = -2.5 \lg{\left( 1 + \frac{\sigma_{\Delta f}}{\Delta f} \right)} \,.
\end{equation}
The user can change the reference values, which can be useful for cases where the ZTF reference catalog gives unsuitable reference magnitudes, e.g., when a reference image contains the considered object around its maximum light.
An example of a difference flux density plot is shown in  Fig.~\ref{fig:antares}.

\subsubsection{External light curves}

Currently, three external photometry data sources are supported: the ANTARES\footref{foot:antares} broker for ZTF alert data, Pan-STARRS DR2\footnote{\url{https://outerspace.stsci.edu/display/PANSTARRS/}} and Gaia DR3\footnote{\url{https://www.cosmos.esa.int/web/gaia/dr3}}. The light-curve plot can include additional epochs from all of them. We used the corresponding APIs to perform a $5\arcsec$ cone search centered on the ZTF object position and choose the closest object to be included in 
the \viewer\ light curve.

\subsubsection{ZTF alert light curve}\label{sec:antares-lc}

\begin{figure*}
    \centering
    \includegraphics[width=\textwidth]{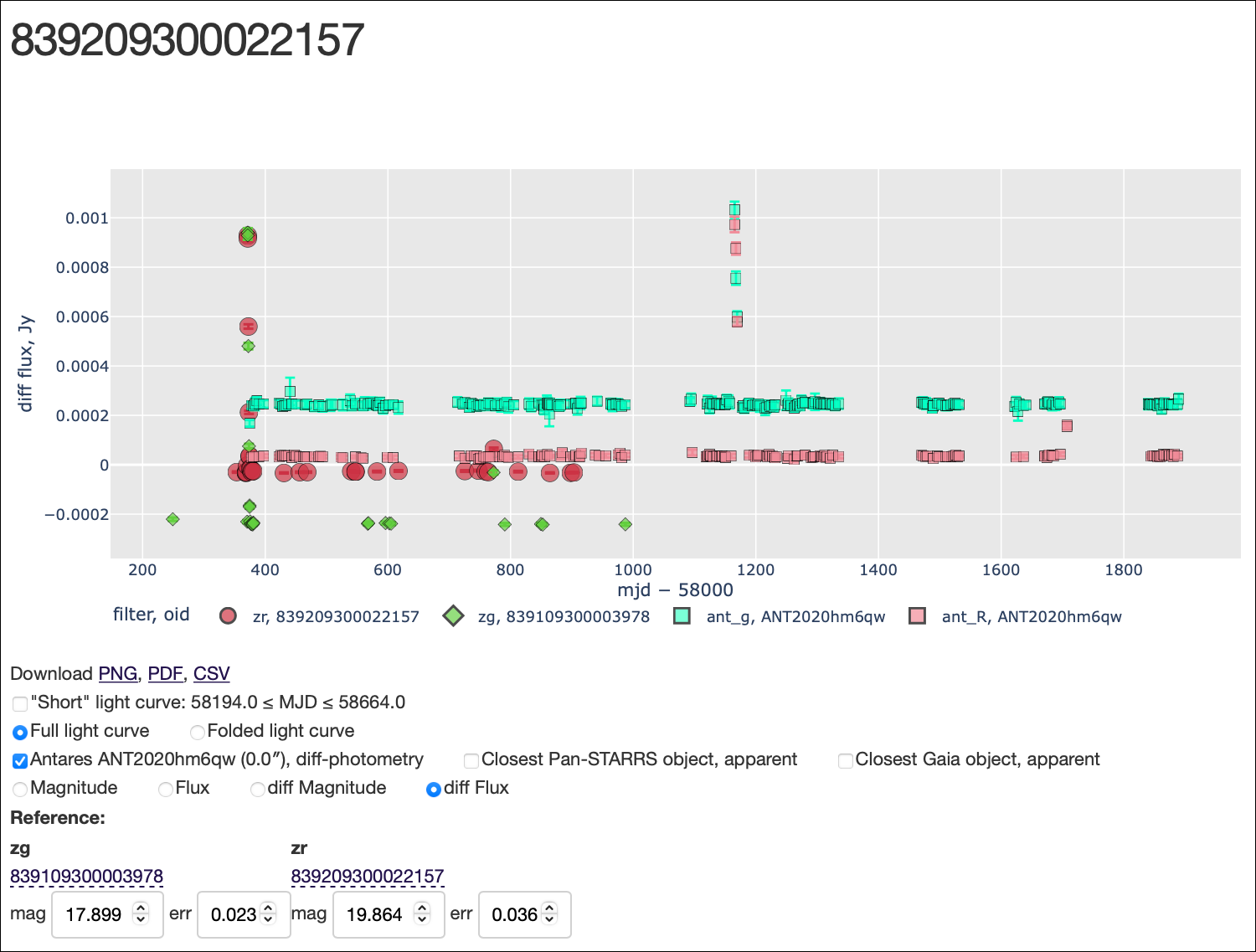}
    \caption{Light curve of a cataclysmic variable \citep{AT2017hue} AT2017hue / ZTF18abvtemh / ZTF DR 839209300022157 in difference fluxes. Alert data (light squares showing the outburst around MJD 59200) alone gives only a limited picture of an object's variability. Here, with ZTF DR4 (dark circles and diamonds with the outburst around MJD 58400) observations, it can be seen that the object is recurrent and has already reached the same level of the brightness before. The bottom part of the screenshot shows the reference magnitude selection block.}
    \label{fig:antares}
\end{figure*}

Apart from the availability of non-detections, ZTF alert pipeline is intrinsically different from those in the DR (see Section~\ref{sec:ztf-drs} and \citet{Bellm_etal2019a} for details).
Moreover, the alert stream delivered to the ZTF brokers may contain more recent observations.
Therefore showing this data together with the ZTF DR light curves might bring important insights on the astronomical objects inspected with the \viewer.
Moreover, ZTF alert users can find the ZTF DR useful to distinguish, for instance, a supernova from an active galactic nucleus or a cataclysmic variable, which could have similar alert light curves, but different DR light curves (see Fig.~\ref{fig:antares} for a real-life example of such a case).
For this purpose we use the ANTARES\footref{foot:antares} broker~\citep{antares} API.\footnote{\url{https://nsf-noirlab.gitlab.io/csdc/antares/client/index.html}}

\subsubsection{Pan-STARRS DR2}\label{sec:pan-starrs-lc}

The \viewer\ also uses data from Pan-STARRS. ZTF uses photometric filters with passbands close to the Pan-STARRS filters, and also uses its data for photometric calibration. Moreover, both surveys operate in the northern hemisphere and cover roughly the same part of the sky. Finally, Pan-STARRS DR2 includes time-resolved photometry data in $grizy$ passbands from a period before the start of ZTF: from 2010 to 2014~\citep{panstarrs}. All these make Pan-STARRS DR2 light curves complementary to ZTF data. Therefore we provide Pan-STARRS DR2 PSF photometry, and we use the MAST API\footnote{\url{https://catalogs.mast.stsci.edu/panstarrs/}} for data access.

Pan-STARRS DR2 gives calibrated flux densities in janskies, which we convert into AB-magnitudes using the inverse of equations \ref{eq:flux_apparent} and \ref{eq:flux_err_apparent},
\begin{equation}
    m = 8.9 - 2.5 \lg{f} \,,
\end{equation}
\begin{equation}
    \sigma_m = \frac{2.5}{\ln{10}} \frac{\sigma_f}{f} \,.
\end{equation}

\begin{figure*}
    \centering
    \includegraphics[width=\textwidth]{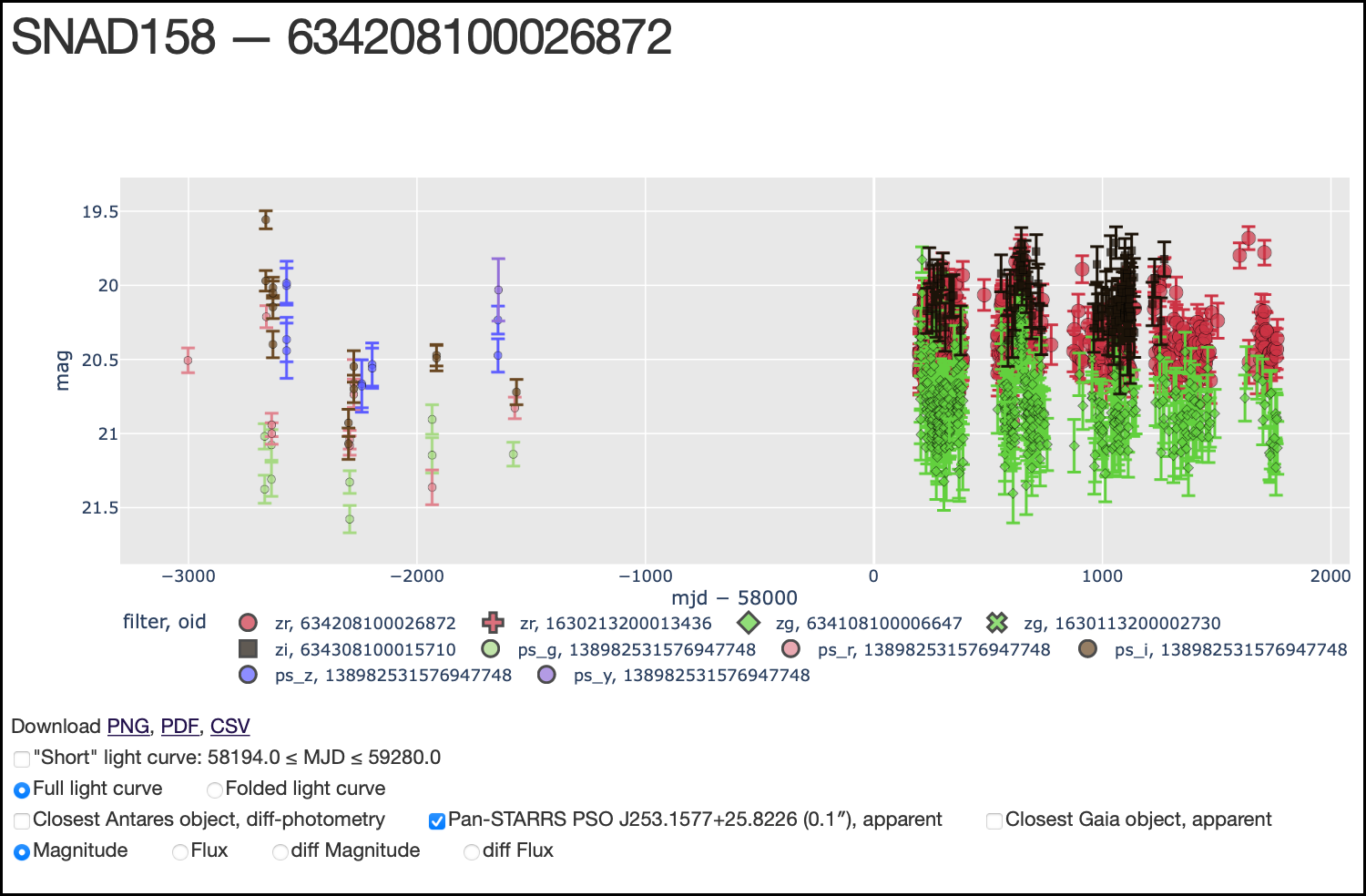}
    \caption{ZTF~DR13 and Pan-STARRS~DR2 light curves of AGN candidate~\citep{Aleo_etal2022} SNAD158 / AT2018lzd / ZTF~DR~OID~\texttt{634208100026872} / PSO~J253.1577+25.8226.}
    \label{fig:pan-starrs-lc}
\end{figure*}

Pan-STARRS provides observation times according to the international atomic time standard at the midpoint of each observation. We transformed this value to  Heliocentric MJD in UTC. An example of a combined ZTF-Pan-STARRS light curve is shown in Fig.~\ref{fig:pan-starrs-lc}.

\subsubsection{Gaia DR3}\label{sec:gaia-lc}

The third data release of Gaia contains epoch photometry in $G$, $BP$ \& $RP$ passbands covering observations between 25 July 2014 and 28 May 2017~\citep{gaiadr3,gaiadr3_variability}.
The Gaia DataLink service\footnote{\url{https://gea.esac.esa.int/data-server/datalink}}~\citep{ivoa_datalink,gaia_datalink} provides access to epoch instrument fluxes,  $\mathfrak{f}_\mathrm{Gaia}$.
We used instrumental zero-points, $\mathrm{zp}$, and their uncertainties, $\sigma_\mathrm{zp}$, to convert fluxes to AB-magnitudes~\citep{gaiadr3_photometry},
\begin{equation}
    m = \mathrm{zp} - 2.5 \lg(\mathfrak{f}_\mathrm{Gaia}) \,,
\end{equation}
\begin{equation}
    \sigma_m = \sqrt{ \left( \frac{2.5}{\ln{10}} \frac{\sigma_{\mathfrak{f}_\mathrm{Gaia}}}{\mathfrak{f}_\mathrm{Gaia}} \right)^2 + \sigma_\mathrm{zp}^2 } \,.
\end{equation}
Flux values and their uncertainties, in janskies, were then obtained using equations \ref{eq:flux_apparent} and \ref{eq:flux_err_apparent}.

\begin{figure*}
    \centering
    \includegraphics[width=\textwidth]{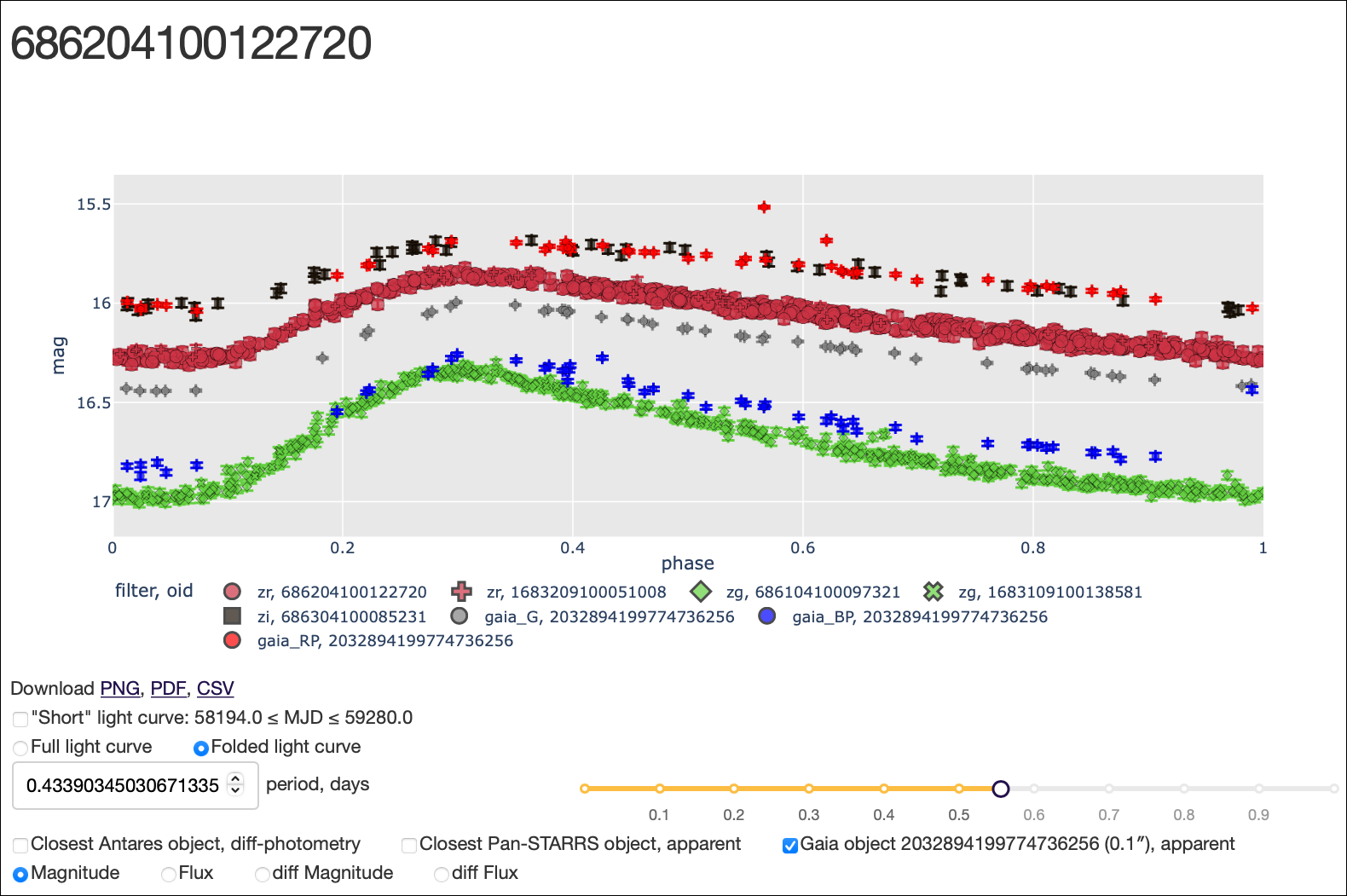}
    \caption{ZTF~DR13 and Gaia~DR3 period-folded light curves of RR Lyrae variable~\citep{Clementini_etal2019} ZTF~DR~OID~\texttt{686204100122720} / Gaia~DR3 Source~ID~\texttt{2032894199774736256}. The period is estimated automatically, while the non-zero phase is selected by the user.}
    \label{fig:gaia-lc}
\end{figure*}

Gaia uses the TCB time standard and provides the average transit Barycentric time.
We converted it to UTC standard and assumed that for the purposes of the majority of the \viewer\ users, the few seconds-level difference between the Barycentric time and the Heliocentric time \citep[e.g.,][]{2010PASP..122..935E} as given in ZTF DR, would not be considered significant.
An example of a combined ZTF--Gaia light curve is shown in Fig.~\ref{fig:gaia-lc}.
 
\subsection{ZTF FITS}\label{sec:js9}

We added an embedded JS9\footnote{\url{https://js9.si.edu}}~\citep{js9} FITS image widget to the right of the light-curve plot (see Fig.~\ref{fig:oid-page}).
The user can click on any ZTF DR photometry point to load the scientific image corresponding to that observation.
The image is centered on the object position, marked by a green dot, and the default zoom factor is equal to unity, which means that a single screen point corresponds to a single CCD pixel. While we used the JS9 widget without the toolbar and the menu, the link to the fully-functional JS9 viewer is provided just below the image itself, in case the user requires additional functionalities.

We used the IRSA IPAC archive\footnote{\url{https://irsa.ipac.caltech.edu/ibe/data/ztf/products/sci/}} as the original source of these FITS images. Additionally, we employed our own proxy to work around the problem of loading data into the JS9 widget from a location not allowing cross-origin resource sharing (CORS)\footnote{\url{https://developer.mozilla.org/en-US/docs/Web/HTTP/CORS}} due to safety concerns (see Section~\ref{sec:fits-proxy} for our implementation details). The archive uses the following path encoding for scientific FITS images:
\texttt{{\color{red}YYYY}/{\color{orange}MM}{\color{purple}DD}/{\color{magenta}MJDFRA}/ztf\_{\color{red}YYYY}{\color{orange}MM}{\color{purple}DD}{\color{magenta}MJDFRA}\_{\color{OliveGreen}FIELDN}\_{\color{brown}PB}\_c} \texttt{{\color{olive}CN}\_o\_q{\color{Cyan}Q}\_sciimg.fits},
where \texttt{\color{red}YYYY}, \texttt{\color{orange}MM}, \texttt{\color{purple}DD} are the year, month, and day of observation in UTC, \texttt{\color{magenta}MJDFRA} is the fractional part of the MJD at the start of exposure,
\texttt{\color{OliveGreen}FIELDN} is the ZTF field number, \texttt{\color{brown}PB} is the ZTF passband name (zg, zr or zi), \texttt{{\color{olive}CN}} is the CCD identifier (01 to 16), and \texttt{\color{Cyan}Q} is the CCD quadrant identifier (1 to 4).
A typical example of a path for a source is:
\texttt{{\color{red}2019}/{\color{orange}03}{\color{purple}14}/{\color{magenta}432616}/ztf\_{\color{red}2019}{\color{orange}03}{\color{purple}14}{\color{magenta}432616}\_{\color{OliveGreen}000633}\_{\color{brown}zr}\_c} \texttt{{\color{olive}07}\_o\_q{\color{Cyan}4}\_sciimg.fits}.

The main issue we faced while assembling this path is that the ZTF DR bulk-downloadable files (Section~\ref{sec:ztf-drs}) do not provide enough information to allow for their unambiguous reconstruction. They provide only the HMJD and the read-out channel identifier \texttt{rcid} (0 to 63).
The latter is sufficient to reconstruct the CCD and quadrant identifiers,
\begin{equation}
    \texttt{\color{olive}CN} = \left\lfloor \frac{\texttt{rcid}}{4} \right\rfloor + 1 \,,
\end{equation}
\begin{equation}
    \texttt{\color{Cyan}Q} = \texttt{rcid} \bmod 4 + 1 \,,
\end{equation}
The more challenging part is reconstructing the MJD of the beginning of the exposure from the HMJD in the middle of it,  without information about the exposure duration and having a precision of around 
a second 
(HMJD is rounded to five decimals).
We solved this issue by getting an estimate of the required \texttt{\color{magenta}MJDFRA}\footnote{We used  \texttt{astropy.time.Time.light\_travel\_time} method to get the difference between MJD and HMJD.} and by identifying the closest available value in the archive, which resulted in a working approximation for the FITS file path.

\subsection{External catalogs cross-matching}\label{sec:external-catalogs}

One of the main purposes of the \viewer\ is to give experts relevant information about a given source beyond what is available within  ZTF DR photometric data.
We achieved this by providing cross-matched information from multiple catalogs, including:
the General Catalog of Variable Stars \citep[GCVS, ][]{gcvs2003,gcvs2017}, the International Variable Star Index \citep[AAVSO VXS, ][]{vsx}, the Atlas Catalog of Variable Stars \cite[ATLAS-VAR, ][]{atlas-var}, the Sloan Digital Sky Survey DR16 Quasar Catalog~\citep{sdssdr16_quasar}, the ZTF Catalog of Periodic Variable Stars~\citep{ztf-periodic}, the Spitzer/IRAC Candidate YSO Catalog~\citep[SPICY, ][]{spicy}, Pan-STARRS DR2~\citep{panstarrs} \added{(stacked photometry table)}, the Transient Name Server\footnote{\url{https://www.wis-tns.org}} (TNS), the Open Supernova Catalog (\citet[OSC, ][]{osc}), the OGLE~III catalog of Variable Stars~\citep{ogle3}, the  SIMBAD astronomical database~\citep{simbad}, the Gaia EDR3 Distance Catalog~\citep{Bailer-Jones_etal2021} and the Gaia DR3~\citep{gaiadr3}.
For most of the listed catalogs, we used
the VizieR service\footnote{\url{https://vizier.u-strasbg.fr}} \citep{vizier},
MAST\footnote{\url{https://archive.stsci.edu}} for Pan-STARRS,
Open Astronomy Catalog API\footnote{\url{https://astrocats.space}} for the OSC,
the ESA Gaia Archive\footnote{\url{https://gea.esac.esa.int/data-server/datalink}},
and our own services for the remaining catalogs (see Section~\ref{sec:catalog-apis} for the implementation details).
For each considered catalog, the \viewer\ displays only a few selected columns,  while also providing a link to the full record at the external resource.
It also provides the cross-match result for the default cone radius but gives the user the ability to set a different value.

\subsection{Light-curve features}\label{sec:light-curve-features}

We created the \viewer\ for the analysis of objects suggested by machine-learning algorithms which use time-series features extracted from ZTF photometrical data.
SNAD team uses the \texttt{light-curve}\footnote{\url{https://github.com/light-curve}}~\citep{Malanchev_etal2021,Malanchev2021} feature extraction toolkit to prepare the input for these algorithms.
Thus, we provide a list of extracted features on the object page of the \viewer, allowing the user to choose between different versions --  
each set corresponds either to a SNAD paper or a major version of the \texttt{light-curve} package.
We also use the highest periodogram peak as the default period for the folded light-curve plot (see Section~\ref{sec:light-curve-plot}) and in the Summary section (see Section~\ref{sec:summary-section}).

\subsection{Summary section}\label{sec:summary-section}

The summary section is an essential part of the object page because it gives the expert a succinct overview of all the data available regarding the chosen source.
Important parts of this section are:

\begin{itemize}
    \item \textbf{Period}: given by light-curve features (see Section~\ref{sec:light-curve-features}) and variable star catalogs.
    \item \textbf{Distance}: collected from all data sources providing a distance or a redshift estimate, the most valuable catalogs for this purpose are Gaia EDR3 distances~\citep{Bailer-Jones_etal2021}, OSC~\citep{osc}, and TNS.
    \item We provide two estimates of \textbf{Extinction}: from the SFD 2-D map~\citep{sfd,schlafly_finkbeiner2011} and the Bayestar 3-D map~\citep{bayestar2015,bayestar2019}, if Gaia EDR3 distance is available. We used the \texttt{dustmaps} Python package~\citep{green2018} in both cases.
    \item Possible object \textbf{types} are collected from all available data sources including catalogs like AAVSO VSX, and live services like TNS.
    \item We also give \textbf{ML classifications} separately to represent probabilistic classification given by ZTF brokers Alerce and Fink, as well as by some catalogs like Gaia~DR3.
    \item \textbf{Average magnitudes and $g-r$ color} are calculated over the current object as well as all selected ZTF neighboring objects.
    \item \textbf{Search in ZTF brokers}: links for cone searches in Alerce, ANTARES, Fink \& Mars.
\end{itemize}

The summary section has a dynamic layout, and therefore it does not show ``missing'' values.
For instance, the  \textbf{distance} and \textbf{type} parts are hidden automatically if no corresponding data are available.
Also, the summary section changes its content on-the-fly if a user changes sections it depends on, e.g., the cone-search radius for some external catalog (see Section~\ref{sec:external-catalogs}) or \texttt{light-curve} version (see Section~\ref{sec:light-curve-features}).

\section{The Knowledge Database}\label{sec:akb}

Beyond the public user-interface features described above, the \viewer\ has a private section called the Knowledge Database (KDB), which is filled and used by authorized experts of the SNAD team.
The KDB was built to allow SNAD experts to share expertise among themselves. Whenever a given candidate is presented to an expert, it is followed by extensive analysis  based on visual screening of all the photometric data in the \viewer,  literature review, photometric model fitting, or follow-up observations.
All this information is combined in a final judgment about the nature of the candidate, which is recorded by the expert in the KDB. 

\begin{figure*}
    \centering
    \includegraphics[width=\textwidth]{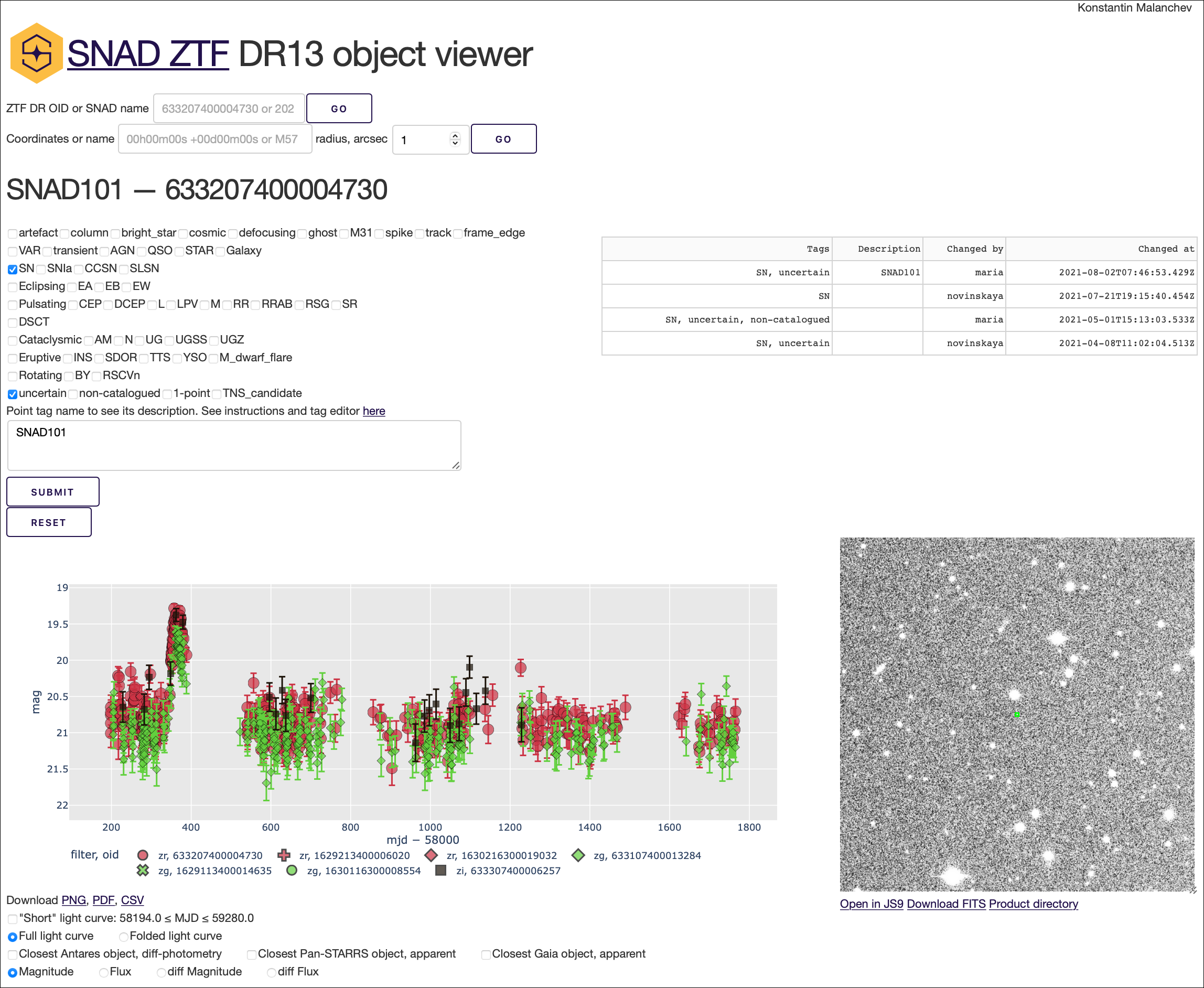}
    \caption{The upper part of the object page contains the SNAD Knowledge Data Base. It represents SN candidate \citep{Pruzhinskaya_etal2022} ZTF~DR~OID~\texttt{633207400004730} / SNAD101 / AT2018lwh. According to this page, the most recent classification of this object is supernova (tag ``SN'') without obvious sub-type (no tags like ``SNIa''), the object does not possess a recorded classification in any of the public catalogs checked by the expert 
    (tag ``uncertain''). The top right table shows that this object was independently inspected by SNAD experts four times.}
    \label{fig:akb-object-page}
\end{figure*}

After login, the expert is directed to an enhanced version of the object page, which contains one more block representing the KDB record for the current OID (Fig.~\ref{fig:akb-object-page}).
The block displays the most recent classification and the description record for the object, as well as a history of all previous records, including its date and author.
The expert can also provide a description of the object in free text format, mainly used to clarify classification choices and extra data sources.

The classification is represented in terms of a non-hierarchical tag system so that each object can have multiple tags. The tag itself is a short string that shows an associated description at pointer hover. Experts can change the order tags and add new ones by accessing the tag editor page. Since the SNAD team works as one coherent group of specialists, every user has the power to change the tag list, which is available to all users. However, the same structure can be personalized for other projects or collaborations, allowing each user to see their own set of chosen tags if necessary. 

Currently, the KDB contains more than 50 tags which can be split into the following categories: types and sub-types of variability such as ``Eclipsing'', ``EA'', ``EB'', and ``EW''\footnote{We adopted types of variability from the GCVS~\citep{gcvs2017}, and VSX~\citep{vsx}: ``Eclispsing'' is used for any eclipsing binaries, while ``EA'', ``EB'' and ``EW'' are sub-types of eclipsing binaries.}; image and photometric pipeline artifacts like ``artifact'' and ``defocusing''; properties of classification such as ``uncertain'' and ``non-cataloged'' which are only meaningful when combined with other tags; and cross-team communication tags like ``TNS\_candidate'', which states the object is worth submitting to TNS.
A detailed description of the currently available tags was reported by \citet{Pruzhinskaya_etal2022}.

\begin{figure*}
    \centering
    \includegraphics[width=\textwidth]{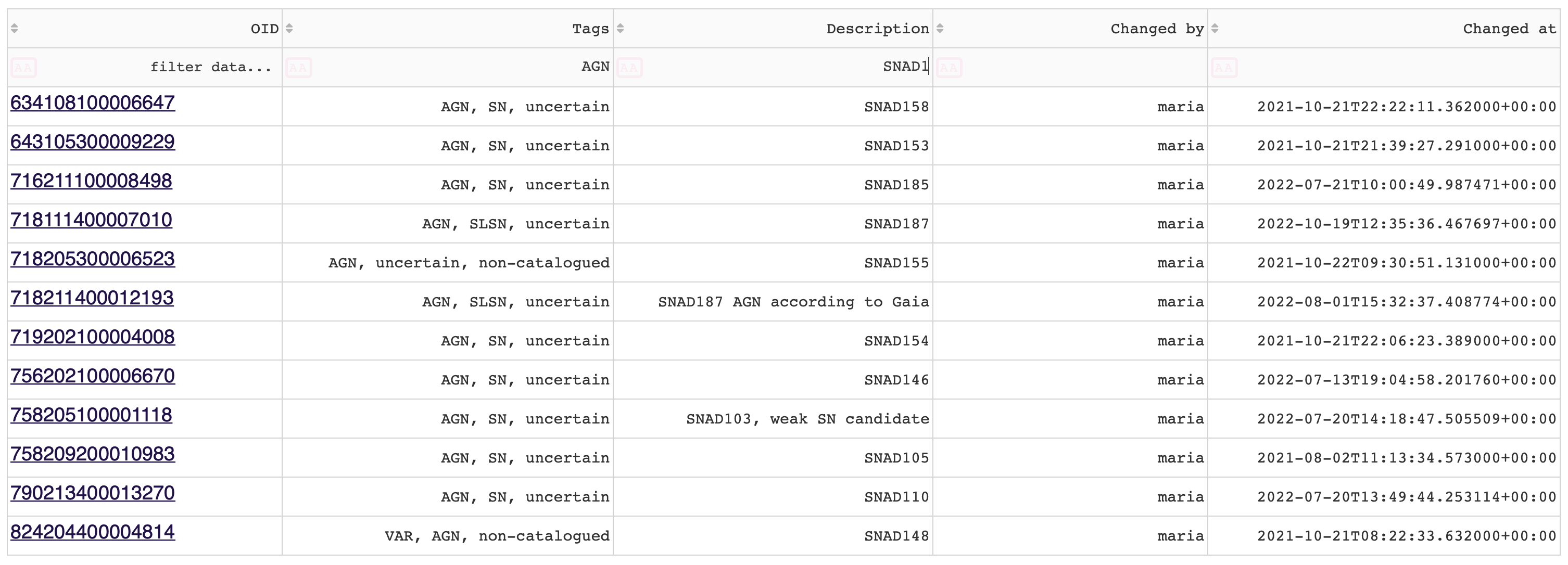}
    \caption{KDB table page. An expert has applied ``AGN'' tag filter and ``SNAD1'' description filter to show AGN candidates from the SNAD catalog.}
    \label{fig:akb-table}
\end{figure*}

Fig.~\ref{fig:akb-table} shows the KDB table page with an interactive table of all tagged objects. An expert can apply filters to individual columns, for example, to see all objects marked with a specific tag or all records modified by someone. The KDB also has programmatic access, which can be used for more sophisticated analysis. We further describe the implementation details of the KDB API service in Section~\ref{sec:akb-api}.

In the context of the adaptive learning algorithms developed by the SNAD team, the final state of a specific learning model is a direct consequence of the feedback provided by the expert during training. 
Thus, the historical record stored in the KDB is crucial to allow for reproducibility of results, as well as to isolate probable causes for divergence in models trained by different experts.
The historical record also allows experts to immediately access each other's judgments, which has proven efficient, especially when experts come from different scientific backgrounds or train models with different goals.
Finally, the SNAD team is currently developing machine learning models which aim at using tags from the KDB as prior information. This will optimize the allocation of human resources since experts will not need to start from scratch when training a model for a new purpose.

\section{Infrastructure}
\label{sec:infrastructure}

The viewer infrastructure is built from multiple services which mainly communicate via the HTTP protocol~\citep{10.17487/RFC2616}. 
We define in this paper a module as a program configured as a single \texttt{Docker}\footnote{\url{https://docker.com}} container, a service as a closely connected set of modules that are configured by a single \texttt{Docker-Compose} file, and a service instance as a set of containers, volumes, and networks deployed by a  \texttt{docker-compose} tool.
Since the \viewer\ does not have high availability requirements, most of our services have only one instance running on a virtual private server (VPS).
However, the ZTF DR database API (see Section~\ref{sec:ztf-dr-db-api}) and the ZTF FITS caching proxy (see Section~\ref{sec:fits-proxy}) services have significant hardware requirements which do not meet our VPS budget. Therefore we have instances running those services at dedicated servers located at two academic institutions (currently at the Sternberg Astronomical Institute and at the University of California, Irvine).
Individually, each of these dedicated servers has lower availability than the VPS due to possible network, power, and hardware issues. 
Nevertheless, we duplicated the services in those two services, increasing the availability at the system level.

The choice of the multi-service architecture behind the \viewer\ design is based not only on budget limits and availability requirements but also on the need for low maintainability effort and easy deployment. 
Most of the services are self-consistent and require no persistent data for deployment to a different host, which is a desirable feature whenever the host server needs to be replaced. 
This is also an advantage in case we decide to perform additional on-demand deployments in the future, dispatching cloud instances automatically for instance.
Issue localization and debugging are also easier in a multi-service architecture because, generally, in the case of unexpected problems, a service will fast-fall, making it easy to identify which service is not available at a given moment.
The same reasoning applies to our choice of having one database management system (DBMS) module per service instead of a single centralized DBMS service holding multiple databases: 1) with a single DBMS instance, any outage or network issue would cause an interruption of all services relying on it, and 2) additional network configuration would be required to help services discover and secure access to this DBMS instance, while, in our approach, \texttt{Docker compose} solves this problem via virtual networks connecting each API module to its DBMS companion.
The \viewer\ service itself is not sensitive to the degradation of other services; for example, if one of our catalog services goes down, the \viewer\ shows that this catalog is not available, but otherwise, it continues to operate as usual. The multi-service approach also allows the use of individual services by multiple projects; for instance, the SNAD collaboration accesses the \viewer\ APIs via Python scripts and notebooks.

In our architecture, individual services are configured as a set of \texttt{Docker} containers bundled to have a common private virtual network and volumes using \texttt{Docker Compose}.
Each of the servers we use has a common infrastructure configured as a dedicated \texttt{Docker Compose} and includes the following modules,
\begin{itemize}
    \item \texttt{nginx-proxy}\footnote{\url{https://github.com/nginx-proxy/nginx-proxy}} is a reverse proxy that gives access to our HTTP services via different domain names (virtual hosting);
    \item \texttt{acme-companion}\footnote{\url{https://github.com/nginx-proxy/acme-companion}} gets and automatically renews TLS certificates via Let's Encrypt authority\footnote{\url{https://letsencrypt.org}}, making it possible to have secure HTTPS access throughout the system;
    \item \texttt{dyndns53}\footnote{\url{https://github.com/hombit/dyndns53}} creates and updates type A CNAME records for service domain names via Amazon Web Services' Route~53 DNS\footnote{\url{https://aws.amazon.com/route53/}}.
\end{itemize}

\added{
The \viewer~ design choices are similar to those the ZTF project team made for their Fritz Asrtronomy Marshal\footref{foot:fritz}.
Its infrastructure is also based on multi-service architecture, services are managed by \texttt{Docker-Compose} and communicate via HTTP RESTful APIs.
The Fritz Marshal also acts as an alert broker via its Kowalski component\footnote{\url{https://github.com/dmitryduev/kowalski}}.
Some of the software and framework choices of Fritz and the \viewer\ also match: both use Python as a main programming language and \texttt{aiohttp}\footnote{\url{https://docs.aiohttp.org}\label{aiohttp}} as a framework for APIs.
However some design and software choices differ, for instance, Fritz uses a document data model via the  \texttt{MongoDB} data management system, while the \viewer\ uses relational model (see Section~\ref{sec:data-storage}).
Both approaches have their pros and cons, and as we mention later, our data model choice was primary based on the requirements coming from the SNAD machine learning pipelines, while the document data model would work faster for the needs of the portal.
}

\begin{figure*}
    \begin{adjustbox}{addcode={\begin{minipage}{\width}}{\caption{%
    Diagram of the service infrastructure. Dashed rectangles represent individual services, circles are web modules and complimentary scripts, cylinders are database management systems, and the display is the Portal module. Parallelograms are external services used by the \viewer. Lines show data flow, double circles mark data receivers. Arrows show data exchange between modules of a single service.
    }\label{fig:services}\end{minipage}},rotate=90,center}
   \includegraphics[width=22cm]{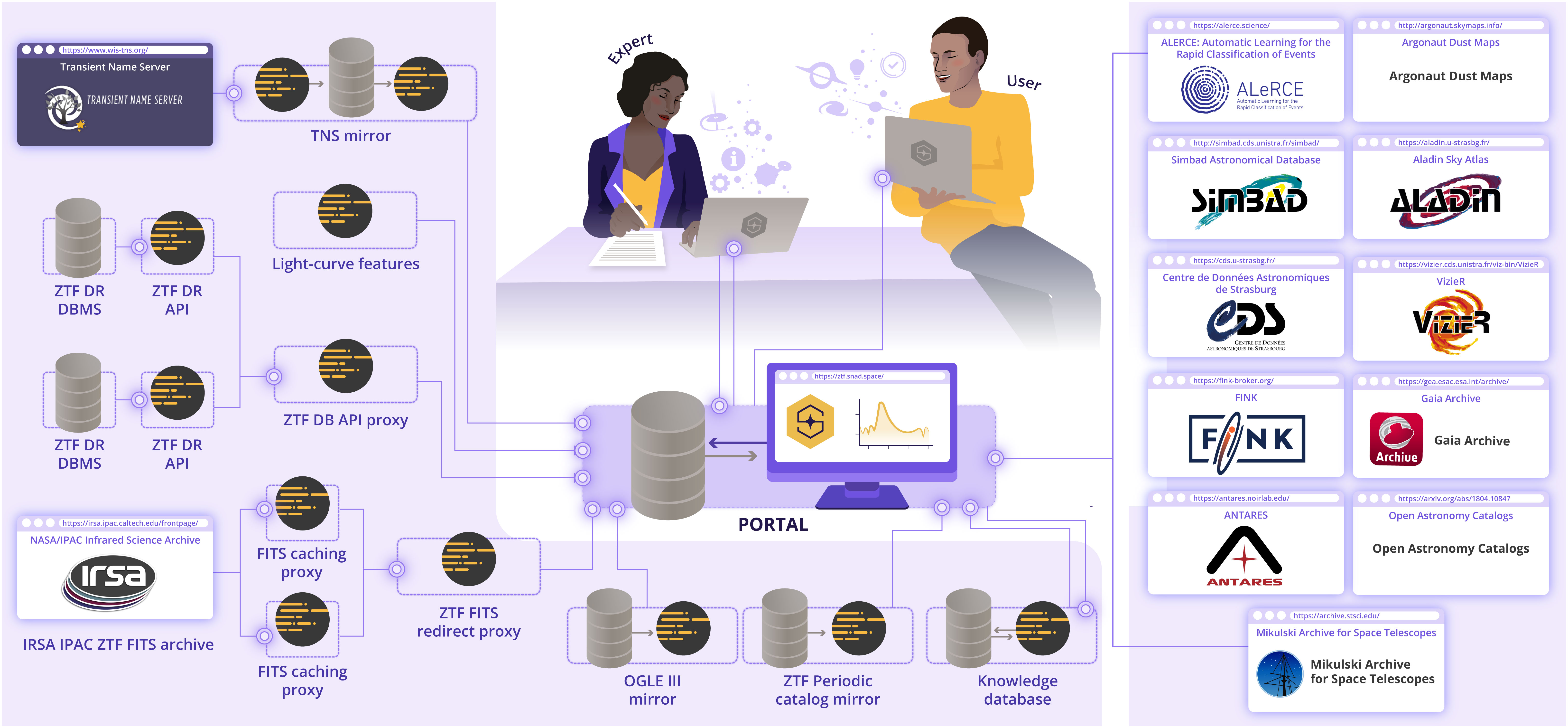}
    \end{adjustbox}
\end{figure*}

We show in Fig.~\ref{fig:services}  the infrastructure of the \viewer, representing an overview of all the services and the individual modules.
The data flow is represented using lines and arrows: the \viewer\ consumes data from many services and external APIs, while experts add records to the Knowledge Database using the \viewer.
We further describe the implementation details of individual \viewer\ services in Sections~\ref{sec:data-storage} and~\ref{sec:impl}.

\subsection{Data storage for ZTF DRs}\label{sec:data-storage}

Since SNAD is focused on the development and application of anomaly detection machine learning algorithms, each of our ZTF DR projects needs to handle millions of light curves containing hundreds of millions of  photometric points.
Therefore one of our main DBMS requirements was a high performance for analyses which query up to a few percent of volume of a multi-terabyte database.
We found that relational columnar databases fit this requirement better than row-based (as \texttt{PostgreSQL}) or document-based (as \texttt{MongoDB}) systems.

Currently the SNAD machine-learning pipeline includes \texttt{Clickhouse} DBMS, which achieved up to $100\times$ higher performance gain over \texttt{PostgreSQL} for our machine-learning needs, while it uses a few times less data storage because of its columnar format and data compression. The \viewer\ employs the same DBMS setup as we use for our machine-learning pipeline, but it uses a very different query pattern: it queries a single ZTF object or a small number of objects for the cone search (Sections~\ref{sec:cone-search-page}~\&~\ref{sec:oid-page}). This usage pattern is sub-optimal for columnar databases, which translates into delays in the \viewer\ usage of ZTF DR DB API, due to \texttt{Clickhouse} response times, up to a few hundred milliseconds.
However, this solution is robust and fits our computation and storage budget since it does not require an additional copy of the database managed by another DBMS.

We use a simple table schema for each DR, which contains detection and object tables and shares a common object ID column used for join queries. Cone-searches are performed using pre-computed H3 indices\footnote{\texttt{Clickhouse} utilises \texttt{H3} library \url{https://h3geo.org}}~\citep[hexagonal hierarchical spherical, ][]{sahr_etal2003} with a resolution of 10, which corresponds to roughly $2.1\arcsec$ tile edge size.
Currently, our \texttt{Clickhouse} database contains ZTF data releases 2, 3, 4, 8, and 13. It has more than a trillion rows and occupies $\sim15$\,TB of storage.

\subsection{Development}\label{sec:development}

The code development follows a multi-service application approach: the source code of each service is located at a separated \texttt{Git}\footnote{\url{https://git-scm.com}} repository; we share no code between the different services. 
The only exceptions are closely related groups of services, namely those coupled with a load-balance service (see Sections~\ref{sec:ztf-dr-db-api},~\ref{sec:fits-proxy}), which, for convenience, use a repository per group. 
We use \texttt{GitHub} as a remote \texttt{Git} repository hosting, issue tracker, collaborative tool via its Pull-request functionality, continuous integration and continuous delivery (CI/CD) tool via its \texttt{GitHub Actions} functionality.

We utilize CI/CD for various types of tests: from the automatic unit, integration, and regression tests for our services to manual user interface tests of the \viewer.
In order to allow for easy manual testing by all the members of the SNAD development team, we use a \texttt{GitHub Actions} workflow which deploys a development version of the \viewer\ for the master \texttt{Git} branch and each active GitHub pull request.
Each such development Portal instance has a domain name \texttt{https://pr\#\#\#.ztf.said.space} with a proper TLS certificate, and runs as a separate \texttt{Docker-Compose} project at a VPS for development.

\section{Services Implementation}\label{sec:impl}
As previously described, the \viewer\ is designed following a multi-service architecture.
In this section, we present its most important services.

\subsection{The Viewer web-portal}

Initially, the \viewer\ was a simple dashboard-like single-page web application for displaying the light curves of ZTF DR1 and a few catalog cross-matches.
The only valuable requirements at that moment were development velocity and usage of the Python language at the back-end, which enabled access to various astronomical packages. At that time, we found \texttt{plotly}\footnote{\url{https://plotly.com}} to be a good solution for interactive graph plotting, allowing us to write both front- and back-ends as a single Python application. We also discovered that the developer of \texttt{plotly} had just released the \texttt{Dash}\footnote{\url{https://dash.plotly.com}} framework, built upon \texttt{Plotly.js} and \texttt{React.js} JavaScript libraries, which allows adding a Python callback to almost every user-interactable HTML tag and which has a set of useful extensions for interactive data representation. Given that \texttt{Dash} fitted our requirements, we decided to implement the \viewer\ in this framework. It is worth mentioning that another useful interactive data visualization framework is available in Python:  \texttt{bokeh}\footnote{\url{https://bokeh.org}}, which also provides tools for graphical data but does not cover control over check-boxes, lists, tables, and other elements of the web-page.

\texttt{Dash} allowed us to rapidly develop an initial version of the portal with a small number of code lines. However, soon after that, we faced some of its limitations. The main issue with the current \viewer\ implementation is limited opportunities for testing (see Section~\ref{sec:development} for more details about the development pipeline) because \texttt{Dash} forces developers to mix data-model and data-view code.
This makes debugging programming issues significantly more challenging and also makes code maintenance more time-consuming.
Another specificity of \texttt{Dash} is the single-page design of the application, which makes it challenging to properly support a rich URL scheme and per-page specification of the tags inside of \texttt{<head>}.\footnote{A limited support of the multi-paging was recently introduced by \texttt{Dash} version 2.}
However, \texttt{Dash} uses the  \texttt{Flask}\footnote{\url{https://flask.palletsprojects.com}} web framework for HTTP management, which allowed us to implement a few non-HTML endpoints (for example, for downloadable plots, see Section~\ref{sec:light-curve-plot}) independently from the main application code.

The source code of the web-portal is available via GitHub\footnote{\url{https://github.com/snad-space/ztf-viewer}}.
The portal wouldn't be possible without the libraries it utilises, including but not limited to:
\texttt{astropy}~\citep{astropy},
\texttt{astroquery}~\citep{astroquery},
\texttt{dustmaps}~\citep{green2018},
\texttt{h5py}~\citep{python_hdf5},
\texttt{healpy}~\citep{Zonca2019},
\texttt{numpy}~\citep{numpy},
\texttt{pandas}~\citep{reback2020pandas,mckinney-proc-scipy-2010},
\texttt{scipy}~\citep{2020SciPy-NMeth}.

\subsection{Catalog APIs}\label{sec:catalog-apis}

This section describes the HTTP API services that we built for the databases we host for cross-matching purposes.
All services are written in Python and use \texttt{Gunicorn} Web Server Gateway Interface (WSGI) HTTP server. Most of the services employ a simple scheme for endpoints, having endpoints like \texttt{/api/v1/circle?ra=RA\&dec=DEC\&radius\_arcsec=RADIUS} typically returning data in JSON format.

\subsubsection{ZTF DR}\label{sec:ztf-dr-db-api}
The ZTF DR database API has three services: 1) \texttt{Clickhouse} DBMS (see Section~\ref{sec:data-storage}), 2) Python API-service which uses a connection to the database and provides HTTP API to access it, and 3) \texttt{Nginx}\footnote{\url{https://nginx.com}} web-server configured to be a fail-over reverse proxy.
The first two services have two instances each and they are continuously running at dedicated servers located at different academic institutions (Section \ref{sec:infrastructure}), while the last one runs at the VPS and proxies queries to the first alive API service instance for higher availability.

The HTTP API service uses the asynchronous Python web framework \texttt{aiohttp}\footref{foot:aiohttp}.
Currently, it gives access to two main \texttt{Clickhouse} tables for each supported ZTF DR, namely the metadata table and the detection table. We have made our source code available on  GitHub\footnote{\url{https://github.com/snad-space/snad-ztf-db}}.

\subsubsection{OGLE III}

Since OGLE~III~\citep{ogle3} provides a user-friendly web interface but does not have a dedicated API, we maintain its mirror.
We used the \texttt{PostgreSQL}\footnote{\url{https://www.postgresql.org}} DBMS and the \texttt{Flask} Python web-framework for the HTTP API implementation. We have made our source code available on  GitHub\footnote{\url{https://github.com/snad-space/snad-ogle3}}.

\subsubsection{ZTF periodic catalog of variable stars}

The set-up for the ZTF periodic catalog of variable stars~\citep{ztf-periodic} mirror is similar to OGLE~III one. This source code is also available on our GitHub\footnote{\url{https://github.com/snad-space/ztf-periodic-catalog-db}}.

\subsubsection{Transient Name Server}

The TNS provides an API that currently implements a 60-second rate limit which is not suitable for our use case, thus also requiring the maintenance of a local mirror from our side. Since TNS is a live service, we need to update its mirror periodically. For this purpose, the TNS mirror service includes not only the \texttt{PostgreSQL} DBMS and the Python \texttt{aiohttp}\footref{foot:aiohttp} modules but also an additional module that periodically downloads the official TNS daily database dump and ingests it into our database. The source code of this service is also available on GitHub\footnote{\url{https://github.com/snad-space/snad-tns}}.

\subsection{Feature extraction API}

The light-curve feature extraction API wraps a few versions of the \texttt{light-curve-feature}\footnote{\url{https://github.com/light-curve/light-curve-feature}} library as a single module written in Rust with the usage of \texttt{Rocket}\footnote{\url{https://rocket.rs}} web-framework. The source code of this service is available on GitHub\footnote{\url{https://github.com/snad-space/web-light-curve-features}}.

\subsection{ZTF FITS caching proxy}\label{sec:fits-proxy}

We utilize \texttt{Nginx} as a caching reverse proxy for the IRSA IPAC archive of ZTF FITS files, mainly due to the safety issue with HTTP Cross-Origin Access politics of web-browsers (see Section~\ref{sec:js9}).
Since we require a client to download the whole FITS image files, whose size is $
\sim$ 38MB each, the available network bandwidth between the client and one of our caching proxies is a limiting factor. Aiming at optimizing both network speed and fault tolerance, we duplicate the \texttt{Nginx} service located on different continents and use \texttt{geo302}\footnote{\url{https://github.com/hombit/geo302}} redirecting proxy which sends a 302 ``Found'' HTTP response containing a URL of the closest alive caching proxy service. The source code is available on  GitHub\footnote{\url{https://github.com/snad-space/ztf-fits-proxy}}.

\subsection{Knowledge Database API}\label{sec:akb-api}

The KDB implementation consists of a Python application written with the \texttt{Django REST framework}\footnote{\url{https://www.django-rest-framework.org}}, while we use the \texttt{PostgreSQL} DBMS module to hold the underlying data.
The REST API has two top-level end-points: one for tags, which have short names, descriptions, and web-page position indexes, and another endpoint for objects, which are identified by ZTF DR OIDs and have descriptions, sets of tags, authorship, and date of the last change. The API also has basic filtering support which, for instance, allows us to request all objects with a specific tag.
Since one of the key KDB requirements is version logging support (see Section~\ref{sec:akb}), we also hold all the history of object states using the \texttt{Django reversion}\footnote{\url{https://pypi.org/project/django-reversion/}} library. The Django application also handles authorization, as we also keep the editor usernames as a part of the stored history. The KDB service is the only service that has user-generated content; therefore this requires backing up the database in the long term. Since the data volume is quite low, of the order of just a few dozen thousand rows, we simply do a daily data dump through the Django interface and upload it to a Google Drive automatically. The source code is available on the GitHub\footnote{\url{https://github.com/snad-space/akb-backend}}.

\section{Conclusions}
\label{sec:conclusions}

This work describes the SNAD \viewer\, which aims to help address one of the most daunting challenges of contemporary science: how to make sense of big data sets generated by modern experiments.
In many fields with strong observational ties, like astronomy, despite undeniable progress in automatic learning techniques, a significant fraction of potential breakthroughs are still expected to require human screening and intervention for the foreseeable future.
We conceived the SNAD \viewer\ to optimize the allocation of human resources in such tasks.

Our initial goal was to centralize external information about the objects present in ZTF data releases through a single web interface, thus allowing SNAD experts to provide feedback to adaptive learning pipelines more efficiently.
However, we evolved the \viewer\ to be a host of our team expertise, and now it also hosts a knowledge database containing thousands of annotations which we expect to provide valuable priors to inform the training and design of future learning algorithms.

Our system is currently based on a multi-service infrastructure, which allowed us to achieve good development speed, simplicity, and the required level of availability. We built it using the \texttt{Dash} framework, giving our users a powerful interactive interface. It is currently highlighted at ``Plotly \& Dash 500'' rating\footnote{\url{https://dash-demo.plotly.host/plotly-dash-500/}}, and it is regularly used from all continents, but Antarctica, serving from a dozen to a few hundred unique visitors per day and responding to hundreds of thousands HTTP requests every month.

The SNAD \viewer\ framework has also proven to be resilient under different user conditions, and it is currently integrated into the ANTARES and Fink brokers, as well as into the Young Supernova Experiment marshal~\citep{Coulter2022_YSEPZ}.
Moreover, it has enabled the development of all the SNAD projects using ZTF data \citep[e.g.][]{Malanchev_etal2021, Aleo_etal2022, Pruzhinskaya_etal2022}, including the discovery of lost transient candidates in ZTF DR\footref{foot:cat}. 

In this era of big data, the SNAD \viewer\ is an illustrative example of the potential enabled by the open science policies of 21$^{\rm st}$ century astronomy. Its development was only possible due to the large efforts allocated to the maintenance of public data archives and APIs, which guarantees accessibility of final data products to the entire astronomical community, and consequently optimizes the scientific results from these data. It is also a statement on the effort necessary to nurture a truly interdisciplinary environment. The development of its features was, from the very beginning, guided by domain experts who voiced their needs and concerns raised during the analysis process. This experience will certainly be valuable once data from LSST becomes available.

Finally, the SNAD \viewer\ framework described in this work was designed to allow easy adaptation as well as scalability to other surveys.
In the era of LSST, such data centralization about specific objects will be as important to most astrophysical domains as it already is today for multi-messenger astronomy. 
However to realize this vision, tools used for different science cases must be designed to be easily adaptable to the requirements of different research sub-fields.
With the SNAD \viewer\ we present an example of such a tool, and its successful experience so far also demonstrates the great potential they hold for the future of astronomical discovery.

\section*{Acknowledgements}

Authors are grateful to Kirill Sokolovsky, Adam Scott, Julien Peloton, and  Vadim Krushinsky for the helpful discussions. We thank all users of the \viewer\ for their feedback and bug reports. We thank Clara Heinrich for her work in the illustration of Figure \ref{fig:services}. 

This research has made use of the NASA/IPAC Infrared Science Archive, which is funded by the National Aeronautics and Space Administration and operated by the California Institute of Technology.
This research has made use of the SIMBAD database, VizieR catalogue, and ``Aladin sky atlas'' access tool, operated at CDS, Strasbourg, France. This work has made use of results from the ESA space mission Gaia, the data from which were processed by the Gaia Data Processing and Analysis Consortium (DPAC).
Funding for the DPAC has been provided by national institutions, in particular the institutions participating in the Gaia Multilateral Agreement.
Some of the authors are members of the Gaia Data Processing and Analysis Consortium (DPAC).

We used the equipment funded by the Lomonosov Moscow State University Program of Development.
This work made use of server hosting services from the Donald Bren school of Information and Computer Sciences of the University of California, Irvine, and of a machine acquired with the UCI 2021-2022 Professional Development Award.
This work made use of the Illinois Campus Cluster, a computing resource that is operated by the Illinois Campus Cluster Program (ICCP) in conjunction with the National Center for Supercomputing Applications (NCSA) and which is supported by funds from the University of Illinois at Urbana-Champaign.

The reported study was funded by RFBR according to the research project 20-02-00779.
M.V.P. acknowledges the support by the Interdisciplinary Scientific and Educational School of Moscow University “Fundamental and Applied Space Research”
E.E.O.I. and E.R. received financial support from CNRS International Emerging Actions under the project \textit{Real-time analysis of astronomical data for the Legacy Survey of Space and Time} during 2021-2022.
P.D.A.\ is supported by the Illinois Survey Science Graduate Fellowship from the Center for AstroPhysical Surveys (CAPS)\footnote{\url{https://caps.ncsa.illinois.edu/}} at the National Center for Supercomputing Applications (NCSA).
A.K.M. acknowledges the support from the Portuguese Funda\c c\~ao para a Ci\^encia e a Tecnologia (FCT) through grants UID/FIS/00099/2019 for CENTRA and EXPL/FIS-AST/1368/2021.
Supported by Nonprofit Foundation for the Development of Science and Education "Intellect".

\bibliography{sample631}{}
\bibliographystyle{aasjournal}



\end{document}